\documentclass[a4paper,11pt]{article}

\usepackage[english]{babel}

\usepackage[a4paper,top=2cm,bottom=2cm,left=3cm,right=3cm,marginparwidth=1.75cm]{geometry}

\usepackage{multirow}
\usepackage{amsmath}
\usepackage{graphicx}
\usepackage[colorlinks=true, allcolors=blue]{hyperref}
\usepackage{csquotes}
\usepackage[backend=biber,style=numeric,sorting=none]{biblatex}
\usepackage{authblk}
\usepackage{cuted}
\usepackage{caption}
\usepackage{algpseudocode}
\usepackage{algorithm}
\usepackage[most]{tcolorbox}
\usepackage{xcolor}
\usepackage{amssymb}

\usepackage{tabularx}
\usepackage{array}
\usepackage{booktabs}

\newcommand{\figref}[1]{%
    \hyperref[{#1}]{\textbf{\textcolor{black}{\ref*{#1}}}}%
}

\providecommand{\keywords}[1]
{
  \small	
  \textbf{\textit{Keywords---}} #1
}

\title{Long-horizon Forecasting of Sodium-ion Battery \\ Charge-profile Evolution and Degradation \\ using Autoregressive Deep Sequence Models}

\author[1,2,$\dagger$]{Xu Zhang}
\author[1,2,$\dagger$]{Florent Magaud}
\author[1,2]{Mahdi Abid}
\author[1,4]{\\Basile Jezequel}
\author[1,2]{Raynald Lesieur}
\author[1,2,3,4,*]{Arnaud Demortière}

\affil[1]{Laboratoire de Réactivité et de Chimie des Solides (LRCS), CNRS UMR 7314, Université UPJV, Hub de l’Energie, Rue Baudelocque, 80039 Amiens Cedex, France.}
\affil[2]{Réseau sur le Stockage Electrochimique de l’Energie (RS2E), CNRS FR 3459, Hub de l’Energie, Rue Baudelocque, 80039 Amiens Cedex, France.}
\affil[3]{ALISTORE-European Research Institute, CNRS FR 3104, Hub de l’Energie, Rue Baudelocque, 80039 Amiens Cedex, France.}
\affil[4]{Predeeption (CNRS Innovation / INRIA Startup Studio), Hub de l’energie, 80000, Amiens, France}
\affil[$\dagger$]{These authors contributed equally to this work.}
\affil[*]{corresponding author: arnaud.demortiere@cnrs.fr}

\begin{document}
\maketitle

\begin{tcolorbox}[colback=blue!6,colframe=gray!40,arc=4pt,boxrule=0.4pt, left=0.1pt, right=0.1pt, top=6pt, bottom=6pt]

\begin{abstract}

Accurate long-horizon prediction of battery ageing remains challenging because degradation is reflected not only in scalar health indicators but also in progressive changes in the shape and duration of electrochemical cycling profiles. Here, we introduce an autoregressive deep sequence-learning framework for forecasting voltage-current profiles and state of health (SOH) evolution in NVPF/hard-carbon sodium-ion batteries. A stacked LSTM model (Batt-LSTM) and a hybrid CNN-LSTM model (Batt-LSTM-C) are trained on multivariate cycling sequences combining voltage, current, and cycle index, enabling future cycling behavior to be predicted iteratively from an initial window of measured cycling data. This strategy allows degradation trajectories to be forecast directly at the electrochemical-curve level rather than through scalar SOH regression alone. To mitigate the mismatch between training and inference inherent to autoregressive rollout, we implement a scheduled-sampling strategy, extending stable forecasting horizons from a few tens of cycles to up to 600 cycles while preserving local waveform fidelity. Model performance is assessed using complementary metrics, including voltage mean absolute error (MAE), global and local dynamic time warping (DTW), and SOH MAE, to evaluate voltage-profile accuracy and capacity-based health prediction. The predicted profiles further enable the extraction of physically relevant ageing descriptors, including capacity, resistance, and energy-based state-of-health indicators. Epistemic uncertainty is quantified using Monte Carlo dropout and deep ensemble approaches, providing estimates of prediction spread at both local profile and global degradation-trajectory scales. Finally, fine-tuning on a second ageing dataset improves transferability and reduces SOH prediction error during rolling-horizon deployment. These results demonstrate that curve-level forecasting provides a richer and more diagnostic representation of battery ageing than conventional scalar-indicator prediction, offering a promising route toward uncertainty-aware, data-driven battery management and predictive diagnostics.

\end{abstract}

\end{tcolorbox}


\newpage
\section{Introduction}

Sodium-ion batteries are emerging as promising alternatives to lithium-ion technologies owing to the natural abundance of sodium-containing precursors, their potential cost advantages, and their suitability for high-power and fast-charging applications \cite{NVPF_TIAMAT}. However, as in lithium-ion systems, repeated cycling progressively induces performance degradation, including capacity loss, impedance growth, and changes in the electrochemical response of the cell. If such degradation is not properly monitored and anticipated, it can compromise lifetime prediction, safety margins, and the reliability of battery-powered systems \cite{battery_failure}. Battery management systems (BMS) are therefore required to estimate battery states and prognostic quantities, including state of charge (SOC), state of health (SOH), and remaining useful life (RUL), and to adapt operating strategies accordingly. Accurate prediction of these quantities remains a major challenge, particularly under long-term ageing conditions where degradation evolves nonlinearly and depends on both operating history and cell chemistry.

Several modelling strategies have been developed to describe and predict battery behavior \cite{temprano2024advanced}. Empirical models, including equivalent circuit models, represent the battery response through a limited number of electrical components \cite{empirical}. These models are computationally efficient and can be combined with Kalman filtering approaches for real-time SOC and SOH estimation in embedded BMS applications \cite{SOC_SOH_kalman, ANDRE201320}. However, their simplified structure limits their ability to capture complex degradation mechanisms and long-term ageing dynamics. Physics-based models, such as Doyle-Fuller-Newman type electrochemical models, provide a more detailed description of transport, reaction, and internal cell processes \cite{Pybamm, Delacourt_2012}. Nevertheless, their computational cost, sensitivity to parameterization, and difficulty in describing long-term ageing under realistic operating conditions limit their direct deployment in BMS environments. Both empirical and physics-based approaches may also lose accuracy when the battery evolves away from the assumptions or parameters used to build the model \cite{data_vs_model}.

Data-driven methods have therefore attracted increasing interest for battery prognosis and health monitoring. Classical machine learning approaches, such as support vector machines, random forests, and autoregressive models, have been applied to SOC, SOH, and RUL estimation \cite{SVM_RUL, RF_RUL, ARMA_RUL}. These methods can provide efficient predictors when suitable features are available, but they often rely on scalar descriptors or handcrafted indicators extracted from charge-discharge curves. As a result, part of the information contained in the full electrochemical response may be lost. In particular, battery degradation is not only reflected in the decrease of capacity or SOH, but also in the progressive deformation of voltage-current profiles, changes in relaxation behavior, shifts in electrochemical signatures, and the evolution of cycle duration. Capturing these curve-level transformations is essential to obtain a richer and more physically informative representation of ageing.

Recent studies have increasingly exploited electrochemical curves as rich inputs for battery health estimation and degradation prediction. Deep neural networks have been used to reconstruct full charging or open-circuit-voltage curves from partial charging segments and to infer SOH from the reconstructed profiles \cite{lin2024charging,hofmann2024transfer}. In parallel, recent degradation-trajectory models, including generative pre-trained transformers and interpretable mixture-of-experts architectures, have shown the potential of learning long-term ageing evolution from early-cycle or field-accessible signals \cite{hu2026batterygpt,imoe2026}. For sodium-ion batteries, partial charging profiles have also been used for multi-target state estimation across temperatures \cite{liu2025sodium}. However, most existing approaches focus either on scalar SOH/RUL estimation, degradation-trajectory prediction, or reconstruction of a single charging/OCV curve.

In contrast, forecasting complete electrochemical cycling profiles offers an alternative strategy to scalar health-indicator prediction. Instead of predicting SOH directly, the model learns the temporal evolution of the voltage and current response, from which capacity, SOH, internal resistance, and degradation trends can be extracted. Such an approach is particularly relevant for laboratory ageing datasets, where charge-discharge profiles are acquired over thousands of cycles under controlled conditions. Learning from these trajectories can provide reference degradation patterns that may later support adaptive BMS strategies, ageing-aware control, and the detection of deviations from expected behavior \cite{EV}. However, electrochemical profile forecasting is challenging because cycling data are multivariate, non-stationary, and characterised by long-range dependencies, evolving cycle duration, and accumulated prediction errors during long-horizon rollout.

Time-series forecasting  \cite{lim2021time} provides a natural framework to address this problem. In multivariate time series, the future evolution of a signal depends not only on its own history, but also on the coupled evolution of other channels. In battery cycling data, voltage, current, cycle index, and ageing state are strongly interdependent. Deep learning models are well suited to capturing such nonlinear temporal relationships \cite{jorge2023time}. Recurrent neural networks (RNNs) were among the first architectures designed for sequential data, but they suffer from vanishing-gradient limitations when long-term dependencies are involved. Long short-term memory networks (LSTMs) mitigate this issue by introducing gated memory cells that selectively store, update, and forget information over time \cite{LSTM}. LSTM-based models have therefore been widely used for time-series forecasting and battery-state estimation tasks \cite{SOC_LSTM, survey_tsf}. Their performance may be improved by combining them with convolutional layers, which extract local temporal patterns and reduce input dimensionality, or with attention-based mechanisms that enhance long-range contextual representation \cite{CNN_LSTM, Attention_LSTM}.

In this work, we propose an autoregressive deep sequence framework for long-horizon forecasting of sodium-ion battery charge-discharge profiles and degradation. Two architectures are investigated: a stacked LSTM model (Batt-LSTM) and a hybrid CNN-LSTM model (Batt-LSTM-C). Both models are trained on multivariate cycling sequences combining voltage, current, and cycle index, and are used to iteratively forecast future electrochemical profiles from early-cycle measurements. To address the mismatch between training and inference in autoregressive prediction, a scheduled-sampling strategy is introduced to progressively expose the model to its own predictions during training. Model performance is evaluated using complementary metrics, including voltage mean absolute error (MAE), global and local dynamic time warping (DTW), and SOH MAE, to assess voltage-profile accuracy and capacity-based health prediction. Finally, uncertainty quantification and transfer learning are explored through Monte Carlo dropout, deep ensembles, and fine-tuning on a second ageing dataset. This study demonstrates that curve-level forecasting can provide a more informative representation of battery ageing than direct scalar SOH regression alone.

\section{Methods}

This section describes the overall forecasting strategy, data preparation workflow, model architectures, and optimization procedure used to predict the long-term evolution of sodium-ion battery cycling profiles.

\begin{figure}[ht]
    \centering
    \includegraphics[width=\textwidth]{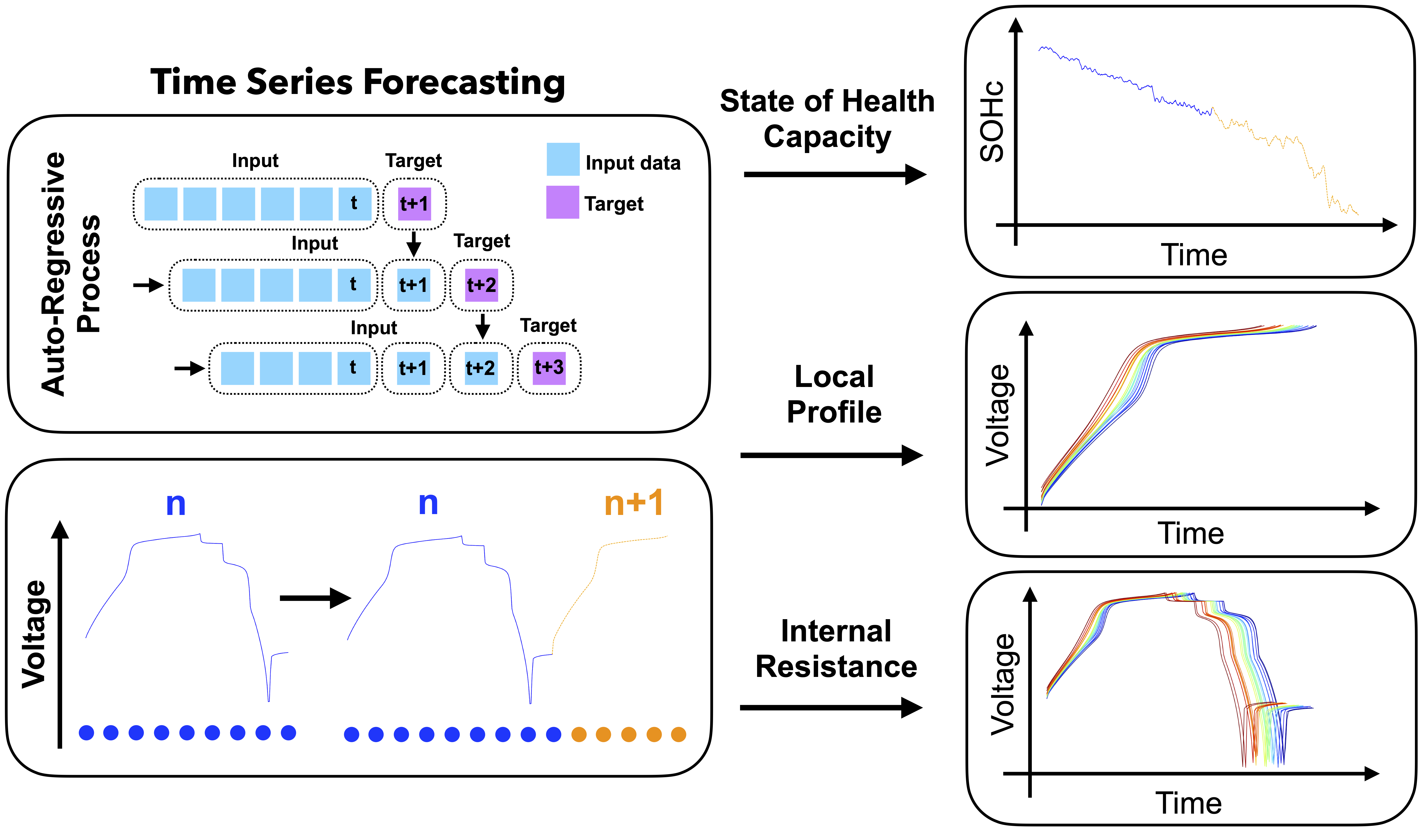}
 %   \captionsetup{font={fontsize=8pt},labelfont={bf,fontsize=8pt}}
 %   \caption{\fontsize{8}{9}\selectfont
    \caption{Schematic overview of the proposed autoregressive forecasting framework for battery ageing prediction. Multivariate cycling sequences are first segmented into input windows and target windows for time-series forecasting. During autoregressive inference, the model predicts the next cycling-profile segment from an initial input window, and the predicted segment is then recursively used as input to forecast subsequent time steps. The resulting long-horizon voltage-current profiles provide access to both local electrochemical profile evolution and global degradation descriptors, including capacity fade, state of health (SOH), and internal-resistance growth.}
    \label{strategy}
\end{figure}

\subsection{Forecasting strategy}

Battery ageing is encoded not only in scalar health indicators, such as SOH or capacity retention, but also in the progressive deformation of the electrochemical cycling profile. The proposed strategy therefore formulates degradation prediction as a curve-level forecasting task, in which the voltage-current response is predicted directly and subsequently used to extract ageing descriptors, including SOH, capacity fade, cycle-duration drift, polarization changes, and internal-resistance trends.

As shown in Figure~\figref{strategy}, the problem is treated as a multivariate autoregressive sequence-to-sequence forecasting task. An input window composed of voltage, current, and cycle index is used to predict the next segment of the cycling profile. The predicted segment is then fed back as the input to the next model call, while the recurrent states are propagated between calls. Voltage and current encode the local electrochemical response, whereas the cycle index provides a global ageing coordinate that supports the prediction of progressive changes in cycle shape and duration.

This strategy enables analysis at both local and global scales. Locally, the model reproduces the evolution of individual voltage profiles, including plateau shifts, relaxation regions, and polarization effects. Globally, the predicted curves are converted into degradation indicators such as SOH, capacity retention, and internal-resistance evolution. Forecasting the full multivariate cycling trajectory therefore preserves richer diagnostic information than direct scalar SOH regression and provides a physically interpretable representation of long-term battery ageing.

\subsection{Extraction of degradation descriptors}

The predicted voltage-current profiles were post-processed to extract global ageing descriptors. In particular, the capacity-based state of health, denoted SOH$_c$, was computed from the cycle-wise capacity as follows:
\begin{equation}
\mathrm{SOH}_c(n) = 100 \times \frac{Q_c(n)}{Q_c(0)} ,
\end{equation}
where $Q_c(n)$ is the capacity estimated from the electrochemical profile at cycle $n$, and $Q_c(0)$ is the initial reference capacity of the cell. 
This definition was applied consistently to both experimental and predicted profiles, enabling direct comparison between ground-truth and forecasted degradation trajectories.

The resistance-based state of health was defined as:
\begin{equation}
\mathrm{SOH}_r(n)= 100 \times \frac{R_{\mathrm{int}}(0)}{R_{\mathrm{int}}(n)},
\end{equation}
where $R_{\mathrm{int}}(n)$ is the internal resistance estimated at cycle $n$. 
With this convention, an increase in internal resistance corresponds to a decrease in SOH$_r$.

The energy-based state of health was computed as:
\begin{equation}
\mathrm{SOH}_e(n) = 100 \times \frac{E(n)}{E(0)} ,
\end{equation}
where $E(n)$ is the energy extracted from the voltage-current profile at cycle $n$, and $E(0)$ is the initial reference energy.

\subsection{Data preparation}

The ageing datasets used in this study were provided by the company TIAMAT \cite{tiamat} and comprise cycling measurements from two independent NVPF/hard-carbon sodium-ion cells cycled under identical laboratory conditions. Both cells were tested at 25~$^\circ$C using a constant-current protocol between 2.0~V and 4.25~V. The cycling sequence consisted of constant-current charge at 2C, followed by a rest period, constant-current discharge at 5C, and a second rest period before the next charge step. Each cell was cycled for approximately 2200 cycles (75 days), providing two long-term ageing trajectories with the same chemistry and operating conditions.

The raw data consist of voltage, current, and time measurements recorded during cycling. Data acquisition was governed by potentiostat-defined logging criteria based on both elapsed time and voltage variation, resulting in a non-uniform sampling grid. To provide regularly sampled inputs to the models used in this study, the raw time series were resampled onto a uniform 15~s time grid using piecewise-linear interpolation. The cycle index was then reconstructed on the interpolated grid and used as an additional input channel. Under constant-current cycling, capacity fade progressively reduces the duration of each cycle. The resulting time series therefore exhibits a non-stationary pseudo-periodicity, in which the cycle length and electrochemical profile evolve continuously with ageing. The final multivariate sequence is therefore composed of three synchronized channels: voltage, current, and cycle index. Each channel was normalized using min--max scaling. The normalization parameters were estimated from the training set only and then applied unchanged to the validation, test, and transfer datasets. The normalization formula is given in Equation~S1.

The continuous multivariate sequence was converted into input--target pairs using a sliding-window strategy. Let $N$ be the total number of time steps, $C$ the number of channels, $L_{\mathrm{in}}$ the input-window length, $L_{\mathrm{out}}$ the number of time steps predicted at each autoregressive step, $s$ the stride, and $N_b$ the number of generated windows. For each window starting at time index $k$, the input sequence is defined as
\begin{equation}
X_k = \left[x_k, x_{k+1}, \ldots, x_{k+L_{\mathrm{in}}-1}\right],
\end{equation}
and the associated target sequence as
\begin{equation}
Y_k = \left[x_{k+L_{\mathrm{in}}}, x_{k+L_{\mathrm{in}}+1}, \ldots, x_{k+L_{\mathrm{in}}+L_{\mathrm{out}}-1}\right],
\end{equation}
where $x_k \in \mathbb{R}^{C}$ contains the voltage, current, and cycle-index values at time step $k$.
This procedure transforms the original sequence of shape $(N,C)$ into a set of input--target pairs of shape $(N_b,L_{\mathrm{in}},C)$ and $(N_b,L_{\mathrm{out}},C)$. A stride of $s=10$ was used to generate overlapping training windows.

The first cell dataset was used for model training, validation, hyperparameter optimization, and in-distribution testing. To construct the test set, one complete cycle was reserved every 100 cycles along the ageing trajectory. During sliding-window generation, training and validation samples were stopped before reaching a reserved test cycle, so that no input--target window could overlap with the test region. The test sequence was then generated by starting from the beginning of the reserved cycle and extending over the predefined number of time steps required for the input and forecast windows. Window generation for training and validation was resumed from the following cycle. This procedure provides test samples initialized at different ageing states while reducing temporal leakage between training, validation, and test data.

\subsection{Autoregressive deep sequence models}

\begin{figure}[ht]
    \centering
    \includegraphics[width=\textwidth]{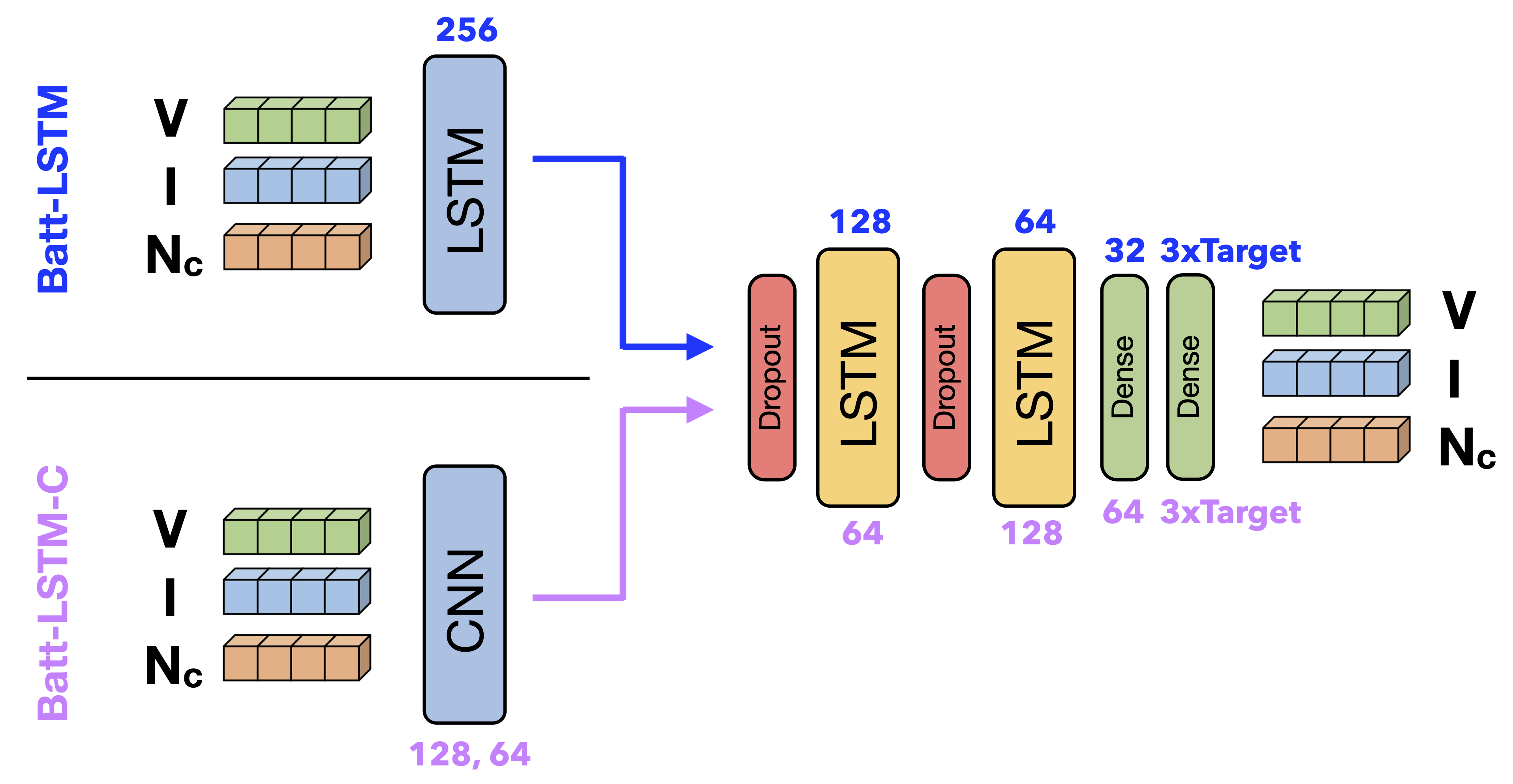}
    \caption{Architectures of the autoregressive deep sequence models used for electrochemical cycling-profile forecasting. The upper branch represents Batt-LSTM, which processes the input sequence through stacked LSTM layers. The lower branch represents Batt-LSTM-C, which applies a convolutional feature-extraction block before the recurrent layers. Dropout is used to regularize the networks, and dense layers map the recurrent representations to the forecast outputs. Blue and purple annotations indicate the layer sizes of Batt-LSTM and Batt-LSTM-C, respectively. The inputs and outputs contain voltage $V$, current $I$, and cycle index $N_c$; ``Target'' denotes the number of time steps predicted per model call.}
    \label{model}
\end{figure}

The forecasting models are designed to predict the future evolution of multivariate electrochemical cycling profiles composed of voltage $V$, current $I$, and cycle index $N_c$. As shown in Figure~\figref{model}, two autoregressive deep sequence architectures were investigated: a stacked LSTM model, referred to as Batt-LSTM, and a hybrid convolutional-recurrent model, referred to as Batt-LSTM-C. Both architectures map an input window of length $L_{\mathrm{in}}$ to the next multivariate sequence segment of length $L_{\mathrm{out}}$. The final dense layer produces $3L_{\mathrm{out}}$ values, which are reshaped into a sequence of shape $(L_{\mathrm{out}},3)$ to jointly predict voltage, current, and cycle index. These predicted segments are subsequently used as inputs during autoregressive rollout.

The Batt-LSTM architecture first encodes the multivariate input sequence using an LSTM layer with 256 units. The resulting temporal representation is then passed through a recurrent prediction head composed of two additional LSTM layers with 128 and 64 units, respectively. Dropout layers are inserted between recurrent layers to reduce overfitting and improve generalization. The final representation is mapped to the forecast window using two fully connected layers, with the last dense layer producing the full multivariate target sequence. The model architecture is summarized in Algorithm~S1 of the Supporting Information.

The Batt-LSTM-C architecture introduces a convolutional feature-extraction block (CNN) before the recurrent prediction head. This block consists of two one-dimensional convolutional layers with 128 and 64 filters, respectively, using a kernel size of 3, a stride of 1, and same padding. Each convolutional layer is followed by max-pooling to reduce the temporal dimension and extract local patterns in the voltage-current profiles. The resulting representation is then processed by the recurrent prediction head, allowing the model to combine local temporal feature extraction with sequence-level memory. The model architecture is summarized in Algorithm~S2 of the Supporting Information.

For both architectures, hyperbolic tangent activation functions are used in the LSTM layers, whereas rectified linear unit activations are used in the convolutional and intermediate dense layers. The final dense layer uses a sigmoid activation because all input and output channels are min-max normalized between 0 and 1. Model predictions are subsequently denormalized for electrochemical analysis and metric evaluation.

The models were implemented in TensorFlow and trained using the Adam optimizer with mean squared error (MSE) as the training loss. Unless otherwise specified, a batch size of 32, a learning rate of $10^{-4}$, and a dropout rate of 0.1 were used. Training was performed on two NVIDIA RTX A6000 GPUs with 48 GB of memory each using CUDA 12.5. The architecture and hyperparameters were selected through a grid-search procedure, including the number of LSTM layers, the number of dense layers, the number of recurrent units, the convolutional filter sizes, and the dropout rate.
 
\subsection{Hyperparameter optimization}

Hyperparameter optimization was performed to identify the input representation, window configuration, and network architecture providing the best compromise between local profile accuracy and long-horizon autoregressive stability. The explored hyperparameter space included the number of input channels, input-window length $L_{\mathrm{in}}$, forecast length per autoregressive step $L_{\mathrm{out}}$, number of recurrent and dense layers, number of units per layer, dropout rate, and stride. The tested values are summarized in \textbf{Table S1}.

Model selection could not rely on mean squared error (MSE) alone, because pointwise errors are sensitive to temporal shifts during autoregressive rollout. A small shift in cycle timing may produce a large MSE even when the predicted electrochemical profile retains a realistic shape. Conversely, a low MSE does not necessarily indicate that the model correctly reproduces the charge--discharge transitions or the evolution of cycle duration. Parameter configurations were therefore assessed using complementary measures of pointwise accuracy, cycle-profile similarity, and temporal drift. MSE was considered together with cycle-wise dynamic time warping (DTW) and differences in the timing of matched complete cycles. Cycle segmentation was based on current transitions, allowing predicted and experimental cycles to be compared. Configurations producing profiles that could not be reliably segmented were not considered suitable for cycle-based degradation analysis. Parameters were selected to balance local waveform fidelity and long-horizon forecasting stability. Figure~S1 illustrates this trade-off for different target-window lengths.

The first optimization step consisted of selecting the most informative input channels. A univariate model trained only on voltage was able to reproduce the local shape of individual cycles, but rapidly converged toward nearly flat profiles during autoregressive rollout. Adding current as a second input channel transformed the task into a multivariate forecasting problem and substantially improved long-horizon stability, because the model could distinguish charge, discharge, and rest periods. However, voltage-current inputs alone still led to an accumulation of temporal drift, mainly due to errors in the predicted cycle duration. To mitigate this effect, the reconstructed cycle index $N_c$ was added as a third input channel. This additional ageing coordinate reduced drift and helped the model capture the progressive shortening and deformation of cycles during degradation. Other candidate channels, including voltage and current derivatives and positional encodings, were evaluated but did not improve the forecasting metrics. The final input representation was therefore composed of voltage $V$, current $I$, and cycle index $N_c$.

The second optimization step focused on the input-window length and the forecast length per autoregressive step. Increasing $L_{\mathrm{in}}$ provided only a moderate improvement in validation metrics while significantly increasing training time. An input length of $L_{\mathrm{in}}=800$ time steps, corresponding to approximately four cycles after resampling, was selected as a compromise between temporal context and computational cost. The choice of $L_{\mathrm{out}}$ was more critical. Short sequence-to-point predictions produced smooth local profiles but accumulated drift rapidly during rollout. In contrast, longer sequence-to-sequence predictions reduced the number of autoregressive iterations and therefore decreased drift, but they also introduced higher waveform noise and degraded local curve fidelity. A forecast length of $L_{\mathrm{out}}=10$ time steps was selected as the best compromise between curve quality and long-horizon stability (Figure~S1).

Finally, the neural architecture was optimized using grid search. For the Batt-LSTM model, the best configuration consisted of three stacked LSTM layers with 256, 128, and 64 units, respectively, followed by a dense layer with 32 units and a final output layer of size $3 L_{\mathrm{out}}$ (Figure~\figref{model} and Algorithm~S1). The factor 3 corresponds to the three predicted channels, namely voltage, current, and cycle index. A dropout rate of 0.1 was used between recurrent layers to limit overfitting. For the Batt-LSTM-C model, the best configuration used two one-dimensional convolutional layers with 128 and 64 filters for local temporal feature extraction, followed by recurrent layers with 64 and 128 units and dense layers with 64 and $3L_{\mathrm{out}}$ units. These optimized configurations were used in the subsequent autoregressive forecasting, scheduled-sampling, uncertainty-quantification, and fine-tuning experiments.

\section{Results and discussion}

The performance of the proposed autoregressive forecasting framework was evaluated at two complementary scales. At the local scale, we assessed the ability of the models to reproduce the evolution of individual voltage profiles, including plateau deformation, relaxation regions, and polarization changes. At the global scale, we evaluated whether the predicted electrochemical profiles could be used to recover the long-term evolution of the capacity-based state of health, denoted SOH$_c$. The following sections first analyze the baseline autoregressive forecasts obtained with the optimized Batt-LSTM and Batt-LSTM-C architectures, and then examine the effect of scheduled sampling on long-horizon stability.

\subsection{Autoregressive profile forecasting}

During inference, the optimized Batt-LSTM and Batt-LSTM-C models were used in a stateful autoregressive mode, as summarized in Algorithm~S3 of the Supporting Information. The initial experimental input window was first used to initialize the LSTM hidden and cell states, which encode the recent cycling history. The predicted multivariate sequence segment was then reinjected as input for the next forecasting step, while the recurrent states were propagated across iterations. This procedure enables the model to retain temporal memory beyond the fixed input-window length and to generate long-horizon forecasts. To limit error accumulation during extended rollout, the LSTM states were periodically reinitialized every 2000 predicted time steps.

\begin{figure}[ht]
    \centering
    \includegraphics[width=\textwidth]{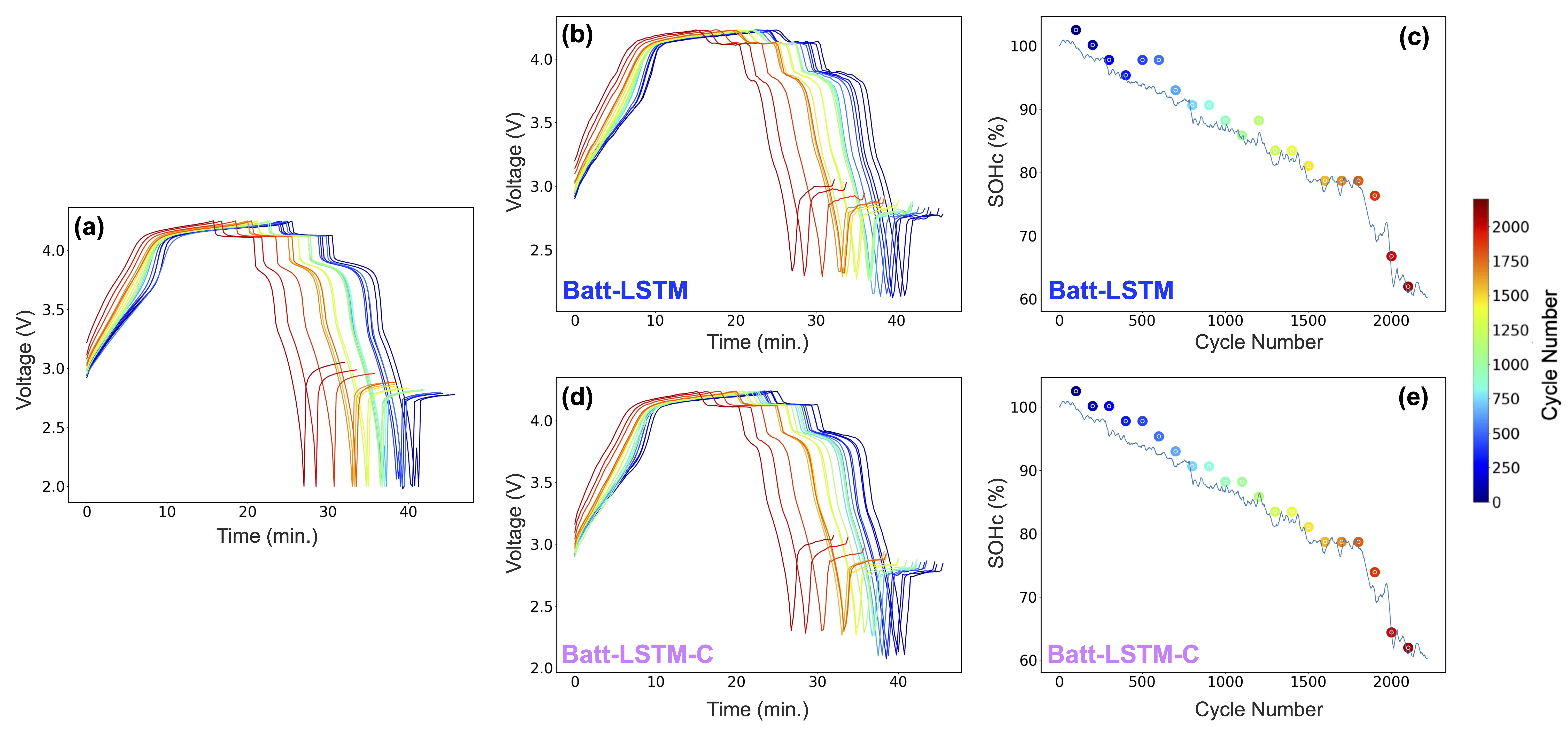}
    \captionof{figure}{Long-horizon autoregressive forecasting of voltage profiles and capacity-based state of health. (a) Stacked experimental voltage profiles used as ground truth, with colors indicating cycle number. (b) Predicted voltage profiles obtained with the Batt-LSTM model and (c) corresponding comparison between the experimental SOH$_c$ trajectory (blue line) and the SOH$_c$ values extracted from the predicted profiles. (d) Predicted voltage profiles obtained with the Batt-LSTM-C model and (e) corresponding SOH$_c$ comparison. The results show that both architectures reproduce the progressive deformation of the charge-discharge profiles and enable extraction of the global SOH$_c$ degradation trend from curve-level forecasts.}
    \label{result}
\end{figure}

Figure \figref{result} compares the experimental voltage profiles with the profiles predicted by the Batt-LSTM and Batt-LSTM-C models. The stacked experimental profiles in Figure \figref{result}a show the progressive evolution of the voltage response during ageing, including the shortening of the cycling duration and the deformation of the charge-discharge profile. Both models reproduce the main local features of the electrochemical curves. In particular, the predicted profiles obtained with Batt-LSTM (Figure \figref{result}b) and Batt-LSTM-C (Figure \figref{result}d) capture the overall voltage evolution, the plateau regions, and the relaxation behavior after current interruption. This indicates that the models learn meaningful local representations of the cycling dynamics rather than only fitting a scalar degradation indicator.

The predicted profiles were then converted into SOH$_c$ values and compared with the experimental SOH$_c$ trajectory (Figure~\figref{result}c,e). Although both architectures provide accurate local profile reconstructions, their long-horizon degradation trajectories remain limited when trained only with the standard sequence-to-sequence objective. This limitation does not primarily arise from an inability to learn the electrochemical profile shape, since the voltage plateaus, relaxation regions, and ageing-induced profile deformation are reasonably reproduced. Instead, it originates from the closed-loop nature of autoregressive inference: prediction errors are progressively fed back into the model input and accumulate over successive rollout steps.

This behavior reveals a training-inference mismatch. During training, the models are mainly exposed to experimental input sequences, whereas during inference they must operate on their own predicted sequences. As a result, small local errors can propagate into temporal drift, inaccurate cycle-duration prediction, and eventual saturation of the SOH$_c$ trajectory toward a quasi-stationary degradation state. Increasing the forecast length per step would reduce the number of autoregressive iterations, but this strategy was shown to degrade local waveform fidelity by introducing noisier voltage profiles. We therefore addressed this limitation at the training level by introducing scheduled sampling, which progressively exposes the model to its own predictions while preserving short-step curve accuracy.

\subsection{Scheduled sampling for long-horizon stability}

Scheduled sampling was implemented to improve the robustness of the models under closed-loop autoregressive inference. The objective is to reduce exposure bias by progressively bridging the gap between the training regime, where ground-truth sequences are available, and the inference regime, where the model must recursively condition on its own predictions \cite{scheduledsampling}. In standard teacher forcing, ground-truth targets are used as inputs during training, which stabilizes optimization but does not prepare the model for error accumulation during rollout. Conversely, free-running training exposes the model to its own predictions, but may lead to unstable convergence at early epochs because the predictions are initially inaccurate.

Scheduled sampling provides an intermediate curriculum between these two regimes. At the beginning of training, the probability of using the experimental sequence as the next input is high, allowing the model to first learn stable local electrochemical profile reconstruction. This probability is then decreased linearly with the epoch number, progressively replacing ground-truth inputs with model-predicted inputs. The model is therefore gradually exposed to the same autoregressive conditions encountered during inference, while avoiding the instability of fully free-running training from the first epochs.

\begin{figure}[ht]
    \centering
    \includegraphics[width=\textwidth]{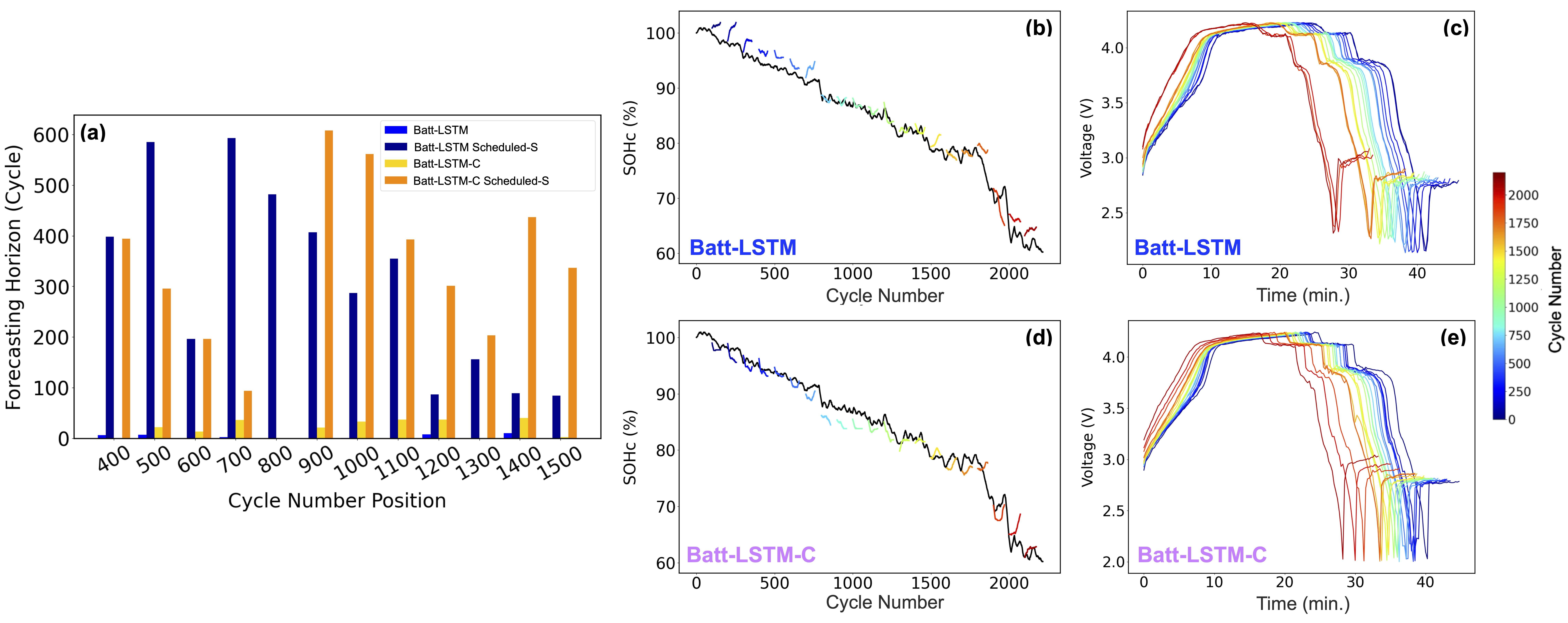}
    \captionof{figure}{Effect of scheduled sampling on long-horizon autoregressive forecasting. (a) Forecasting horizon obtained for Batt-LSTM and Batt-LSTM-C with and without scheduled sampling, using an SOH$_c$ error threshold of 5 percentage points. (b,d) SOH$_c$ trajectories predicted with scheduled sampling for Batt-LSTM and Batt-LSTM-C, respectively, compared with the experimental degradation curve. (c,e) Corresponding short-horizon voltage profiles predicted by Batt-LSTM and Batt-LSTM-C. Scheduled sampling substantially increases the stable forecasting horizon while preserving local voltage-profile fidelity.}
    \label{scheduled_sampling}
\end{figure}

In this work, the probability of using the experimental sequence was set high at the beginning of training and decreased linearly with the epoch number. At each training iteration, the next input segment was therefore selected either from the ground truth or from the model prediction. This curriculum allows the model to first learn stable local profile reconstruction and then progressively adapt to the autoregressive conditions encountered during inference.

The impact of scheduled sampling is summarized in Figure \figref{scheduled_sampling}. Without scheduled sampling, both Batt-LSTM and Batt-LSTM-C exhibit short stable forecasting horizons, typically limited to a few tens of cycles. After scheduled-sampling training, the stable horizon increases substantially, reaching up to approximately 600 cycles depending on the starting point along the ageing trajectory (Figure \figref{scheduled_sampling}a). This improvement demonstrates that the main limitation of the baseline models is not their ability to encode local electrochemical profiles, but rather their robustness to accumulated autoregressive errors.

The SOH$_c$ trajectories obtained with scheduled sampling are shown in Figure \figref{scheduled_sampling}b,d. Compared with the baseline forecasts in Figure \figref{result}c,e, the scheduled-sampling models follow the experimental degradation trend more closely and delay the onset of divergence during long-horizon rollout. The examples in Figure~\figref{scheduled_sampling}c,e show that the main voltage-profile features are retained after scheduled-sampling training. The short-horizon voltage profiles predicted by Batt-LSTM and Batt-LSTM-C after scheduled-sampling training (Figure \figref{scheduled_sampling}c,e) retain the main electrochemical features of the experimental curves, including the voltage plateaus, relaxation regions, and progressive profile deformation.

Overall, scheduled sampling substantially improves the consistency between training and inference in the autoregressive forecasting framework. By exposing the models to their own predictions during training, it reduces error accumulation during rollout and enables more reliable extraction of SOH$_c$ from predicted voltage-current profiles. These results support the relevance of curve-level autoregressive forecasting for battery ageing analysis, where both local electrochemical profile fidelity and global degradation-trajectory consistency are required.

\subsection{Benchmarking of autoregressive forecasting performance}

To compare all models over a common evaluation window, the maximum benchmark horizon was set to 20\,000 time steps, approximately corresponding to the shortest stable forecasting horizons observed for the scheduled-sampling models across the evaluated starting points (Figure~\figref{scheduled_sampling}a). The horizon of up to 600 cycles reported above represents the maximum stable forecasting horizon achieved at particular starting points along the ageing trajectory.

The six neural model variants were compared with decision tree \cite{steinberg2009cart}, random forest \cite{breiman2001random}, and XGBoost \cite{chen2016xgboost}. FR, TF, and SS denote free-running, teacher-forcing, and scheduled-sampling training, respectively. Table~\ref{tab:final_forecasting_metrics} reports the results at forecast horizons of 1000, 2500, 5000, 10\,000, and 20\,000 time steps.

Performance was evaluated using voltage MAE, global DTW, SOH MAE, and local DTW. MAE measures the mean absolute error between normalized predicted and measured voltages. Global DTW uses the full forecast, whereas local DTW uses matched complete cycles, from charge start to discharge end. Both global and local DTW are computed from normalized voltage sequences using the \texttt{tslearn} Sakoe--Chiba constraint, with a radius of 5 for global DTW and a radius of 0 for local DTW (Algorithm~S4). SOH MAE is expressed in percentage points, using the measured Cycle~0 discharge capacity as the shared reference. MAE and global DTW are averaged equally over all samples. SOH MAE and local DTW are first averaged over matched cycles within each sample, then equally over valid samples. The number of valid samples can differ between models and horizons.

\begin{table}[!htbp]
\centering
\footnotesize
\setlength{\tabcolsep}{3pt}
\renewcommand{\arraystretch}{1.05}

\makebox[\textwidth][c]{%
\begin{tabular}{|l|l|r|r|r|r|r|r|r|r|r|}
\hline
 & & \multicolumn{3}{c|}{Batt-LSTM}
   & \multicolumn{3}{c|}{Batt-LSTM-C}
   & & & \\
\cline{3-8}
Horizon & Metric
 & FR & TF & SS
 & FR & TF & SS
 & \shortstack{Decision\\tree}
 & \shortstack{Random\\forest}
 & XGBoost \\
\hline

1000
 & MAE
 & \textbf{0.020} & 0.031 & 0.021
 & 0.137 & 0.102 & 0.050
 & 0.042 & 0.034 & 0.115 \\
 & DTW global
 & 0.389 & 0.912 & \textbf{0.377}
 & 5.722 & 4.145 & 1.542
 & 1.115 & 0.914 & 4.394 \\
 & SOH MAE
 & 1.105 & 1.289 & \textbf{1.053}
 & 2.287 & 1.330 & 1.245
 & 2.585 & 2.002 & 10.973 \\
 & DTW local
 & 0.184 & \textbf{0.139} & 0.165
 & 0.189 & 0.184 & 0.163
 & 0.208 & 0.140 & 0.508 \\
\hline

2500
 & MAE
 & \textbf{0.058} & 0.104 & \textbf{0.058}
 & 0.263 & 0.238 & 0.127
 & 0.082 & 0.071 & 0.187 \\
 & DTW global
 & 3.218 & 6.888 & \textbf{3.085}
 & 16.303 & 15.160 & 7.862
 & 4.494 & 3.904 & 11.604 \\
 & SOH MAE
 & 1.410 & 2.344 & \textbf{1.253}
 & 2.097 & 1.751 & 1.511
 & 3.107 & 1.976 & 11.875 \\
 & DTW local
 & 0.198 & \textbf{0.144} & 0.173
 & 0.192 & 0.189 & 0.169
 & 0.228 & 0.152 & 0.528 \\
\hline

5000
 & MAE
 & 0.128 & 0.186 & \textbf{0.103}
 & 0.286 & 0.268 & 0.214
 & 0.116 & 0.115 & 0.213 \\
 & DTW global
 & 11.645 & 17.306 & \textbf{8.454}
 & 24.363 & 23.524 & 18.822
 & 9.508 & 9.691 & 18.839 \\
 & SOH MAE
 & 1.590 & 3.649 & \textbf{1.298}
 & 2.838 & 2.103 & 1.559
 & 3.025 & 2.008 & 11.565 \\
 & DTW local
 & 0.205 & \textbf{0.145} & 0.174
 & 0.192 & 0.199 & 0.171
 & 0.224 & 0.153 & 0.538 \\
\hline

10000
 & MAE
 & 0.196 & 0.212 & \textbf{0.166}
 & 0.266 & 0.266 & 0.250
 & 0.185 & 0.172 & 0.232 \\
 & DTW global
 & 25.103 & 27.247 & 21.569
 & 32.746 & 32.973 & 31.321
 & 23.105 & \textbf{21.338} & 29.090 \\
 & SOH MAE
 & 1.711 & 5.560 & \textbf{1.640}
 & 4.445 & 2.272 & 1.668
 & 2.950 & 2.064 & 11.765 \\
 & DTW local
 & 0.203 & \textbf{0.147} & 0.171
 & 0.190 & 0.205 & 0.174
 & 0.214 & 0.153 & 0.539 \\
\hline

20000
 & MAE
 & 0.232 & 0.237 & \textbf{0.215}
 & 0.261 & 0.262 & 0.258
 & 0.226 & 0.227 & 0.247 \\
 & DTW global
 & 41.629 & 42.653 & \textbf{39.526}
 & 45.754 & 45.934 & 45.344
 & 40.014 & 40.801 & 43.599 \\
 & SOH MAE
 & \textbf{1.949} & 7.162 & 2.416
 & 6.331 & 2.481 & 2.206
 & 3.118 & 2.251 & 11.687 \\
 & DTW local
 & 0.199 & \textbf{0.144} & 0.162
 & 0.190 & 0.205 & 0.178
 & 0.213 & 0.152 & 0.548 \\
\hline
\end{tabular}%
}

\caption{Benchmark results at five forecast horizons (in time steps). Values are rounded to three decimal places. Bold values indicate the lowest mean for each metric at each horizon, determined before rounding.}
\label{tab:final_forecasting_metrics}
\end{table}

Batt-LSTM-FR gives the lowest voltage MAE at 1000 and 2500 steps. Batt-LSTM-SS gives the lowest voltage MAE at 5000, 10\,000, and 20\,000 steps and the lowest global DTW at 1000, 2500, 5000, and 20\,000 steps. It also gives the lowest SOH MAE from 1000 to 10\,000 steps. At 5000 steps, Batt-LSTM-SS achieves a normalized voltage MAE of 0.103, a global DTW of 8.454, and an SOH MAE of 1.298 percentage points. At 20\,000 steps, Batt-LSTM-FR gives the lowest SOH MAE, with a value of 1.949 percentage points. Batt-LSTM-TF gives the lowest local DTW at all five horizons, showing that the ranking depends on the metric.

The classical models remain competitive. At 10\,000 steps, random forest gives the lowest global DTW, with a value of 21.338 compared with 21.569 for Batt-LSTM-SS. At 20\,000 steps, decision tree achieves a global DTW of 40.014, close to the lowest value of 39.526 obtained with Batt-LSTM-SS, while random forest has a lower SOH MAE than Batt-LSTM-SS (2.251 versus 2.416 percentage points). XGBoost gives the highest SOH MAE at every horizon in this comparison. The Batt-LSTM-C variants have higher voltage MAE and global DTW than their corresponding Batt-LSTM variants at all five horizons. Overall, Batt-LSTM-SS achieves the lowest voltage MAE at all evaluated horizons from 5000 to 20\,000 time steps and the lowest global DTW at four of the five horizons, supporting the effectiveness of scheduled sampling for long-horizon voltage-profile forecasting.

\subsection{Local and global uncertainty}

Reliable long-horizon forecasting requires both accurate predictions and an estimate of predictive uncertainty. Uncertainty is commonly divided into aleatory uncertainty, associated with irreducible data noise, and epistemic uncertainty, associated with limited data and model uncertainty~\cite{aleatory_epistemic}. Here, we focus on epistemic uncertainty in a setting where only two long-term ageing datasets are available.

Local uncertainty in the predicted voltage profiles was estimated using Monte Carlo dropout~\cite{MC_Dropout}. Dropout layers were kept active during inference, turning the deterministic Batt-LSTM into a stochastic predictor. For each input sequence, 1000 stochastic forward passes were performed with a dropout rate of 0.1. The predictive mean was used as the autoregressive input for the next step, while the standard deviation provided a local uncertainty estimate.

As shown in Figure~S2a,b, the standard Batt-LSTM model gives an average voltage standard deviation of approximately 0.11~V, whereas the scheduled-sampling model gives approximately 0.01~V. The narrower uncertainty band indicates a smaller spread among the stochastic voltage predictions in the illustrated example. This spread describes the sensitivity of the predictions to dropout and should be considered alongside their agreement with the experimental profiles.

Global uncertainty was then evaluated on the capacity-based state of health, SOH$_c$. Because feeding back the predictive mean does not propagate the full range of possible trajectories through the autoregressive forecast, a deep ensemble approach was also used~\cite{survey_uncertainty}. Several Batt-LSTM models were trained with different random initializations, and each generated a complete autoregressive forecast from which an SOH$_c$ trajectory was extracted.

Figure~S2c compares the experimental SOH$_c$ trajectory with the ensemble predictions and their uncertainty band. The average standard deviation was approximately 1.45 percentage points. For the scheduled-sampling configuration, greater variability between independently trained models limited the stability of the ensemble estimate. SOH$_c$ uncertainty was therefore estimated using repeated MC-dropout autoregressive rollouts, with a reported uncertainty below 1 percentage point (Figure~S2d). These results illustrate the prediction spread at both the voltage-profile and SOH$_c$ scales. However, because different estimators were used for global uncertainty, the difference between their uncertainty bands cannot be attributed to scheduled sampling alone.

\subsection{Rolling-horizon fine-tuning for adaptive degradation forecasting}

To evaluate the transferability of the proposed framework, Batt-LSTM trained with scheduled sampling was tested on the second ageing dataset. This dataset was acquired on an independent NVPF/hard-carbon sodium-ion cell cycled under the same operating conditions as the training cell, but with its own degradation trajectory. Such a setting is representative of a practical deployment scenario, where a model trained on one cell must be adapted to another cell with similar chemistry but different ageing kinetics. Recent studies have shown that transfer learning and domain adaptation can substantially reduce the amount of target-cell data required for battery SOH estimation, especially when degradation trajectories differ between cells or manufacturers~\cite{lu2023deep,huang2023fine}.

A first zero-shot transfer experiment was performed by directly applying the pre-trained Batt-LSTM model to Dataset~2, without any additional training on the target cell. The model produced a physically meaningful degradation trajectory, confirming that the features learned from Dataset~1 capture transferable information on voltage-current profile evolution. However, the average SOH$_c$ error remained significant, reaching 12.85 percentage points. This discrepancy is attributed to cell-to-cell variability, including differences in initial capacity, capacity-fade rate, and the onset of the knee region. Although both cells share the same chemistry and cycling protocol, their ageing trajectories are not identical, which limits direct zero-shot generalization.

\begin{figure}[ht]
    \centering
    \includegraphics[width=\textwidth]{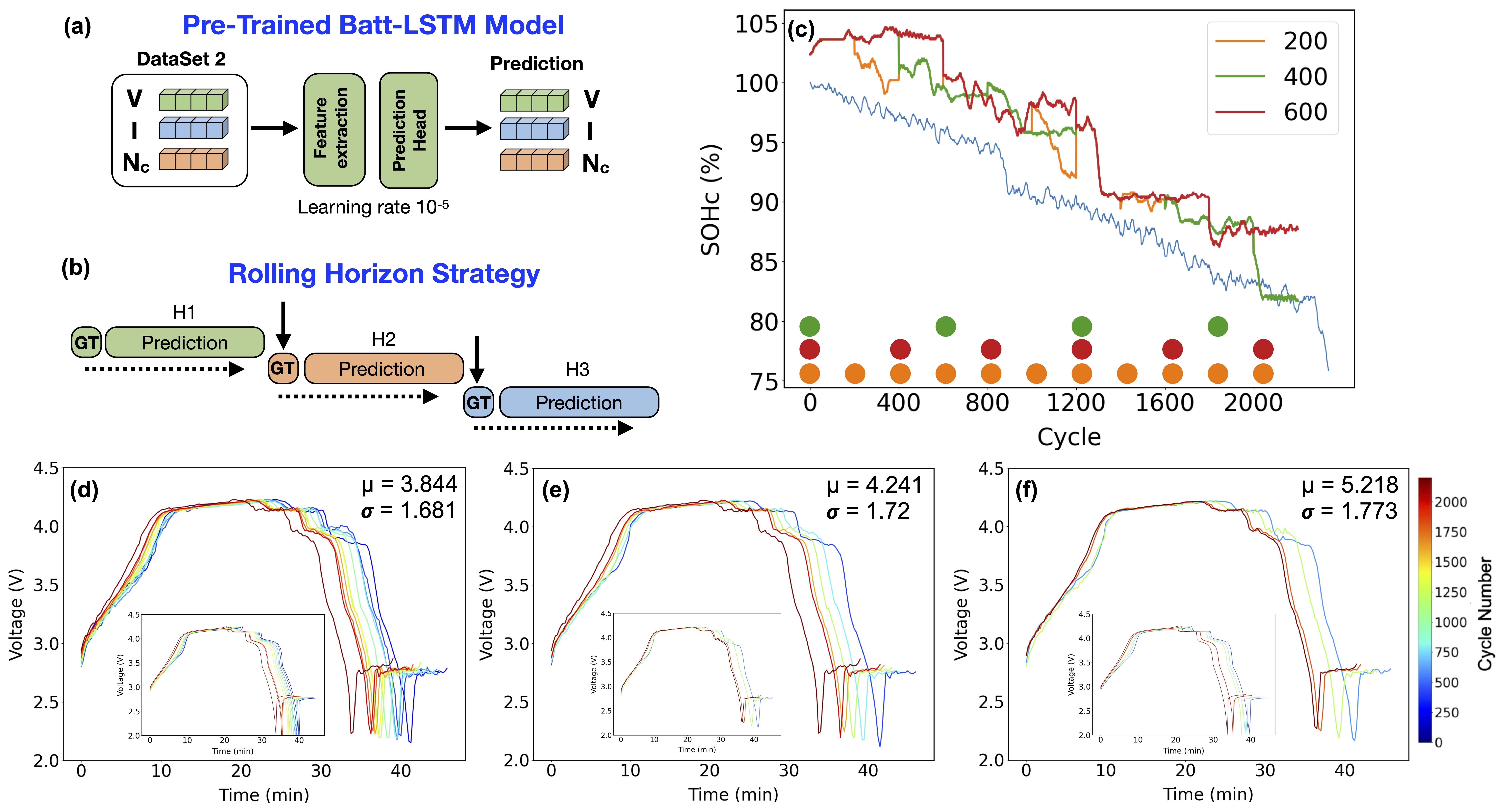}
    \caption{Transfer learning and rolling-horizon fine-tuning on the second ageing dataset. (a) Schematic representation of the fine-tuning strategy applied to the pre-trained Batt-LSTM model using Dataset~2, with voltage $V$, current $I$, and cycle index $N_c$ as input channels. (b) Rolling-horizon forecasting strategy, in which the model is periodically updated with newly available ground-truth data before predicting the next horizon. (c) SOH$_c$ trajectories predicted for rolling horizons of 200, 400, and 600 cycles, compared with the experimental SOH$_c$ evolution. (d-f) Corresponding voltage-profile forecasts obtained for rolling horizons of 200, 400, and 600 cycles, respectively. The mean SOH$_c$ error $\mu$ and standard deviation $\sigma$ indicate that shorter rolling horizons improve degradation tracking and preserve local voltage-profile fidelity.}
    \label{fine_tuning}
\end{figure}

To reduce this domain gap, few-shot fine tuning was performed using increasing amounts of early-cycle data from Dataset~2. As illustrated in Figure~\figref{fine_tuning}a, the pre-trained Batt-LSTM model was adapted to the target cell using voltage $V$, current $I$, and cycle index $N_c$ as multivariate inputs. Three fine-tuning strategies were compared. In the first strategy, all model weights were unfrozen and updated using a reduced learning rate of $1 \times 10^{-5}$. In the second strategy, the recurrent LSTM layers were frozen and only the fully connected prediction head was retrained. In the third strategy, the LSTM layers were frozen while the fully connected layers were reinitialized and retrained. The first strategy, corresponding to full-model fine tuning with a low learning rate, provided the best compromise between preserving the temporal representations learned from Dataset~1 and adapting the model to the degradation kinetics of Dataset~2.

Although fine tuning reduced the SOH$_c$ error, a residual gap remained between the predicted and experimental degradation trajectories. To further improve long-horizon deployment, a rolling-horizon strategy was introduced, as shown in Figure~\figref{fine_tuning}b. In this approach, the model is periodically updated with newly available ground-truth data from the target cell before forecasting the next horizon. This procedure mimics an adaptive monitoring scenario in which the model is progressively recalibrated as new cycling data become available, thereby limiting the accumulation of model bias over long ageing trajectories.

The effect of the rolling-horizon length is shown in Figure~\figref{fine_tuning}c--f. Shorter update intervals provide more frequent correction of the model and therefore improve SOH$_c$ tracking. For a rolling horizon of 200 cycles, the average SOH$_c$ error decreases to 3.844 percentage points, compared with 4.241 and 5.218 percentage points for horizons of 400 and 600 cycles, respectively. The corresponding voltage-profile forecasts, shown in Figure~\figref{fine_tuning}d--f, confirm that shorter horizons also help preserve local electrochemical profile fidelity, particularly near the end of life where degradation accelerates. These results indicate that fine tuning combined with rolling-horizon recalibration provides a practical strategy for adapting a pre-trained autoregressive degradation model to a new cell while limiting the amount of target-cell data required.

The main limitation of this strategy is the need for periodic access to new ground-truth cycling data for model updating. Nevertheless, in a battery monitoring context, this adaptive recalibration is consistent with practical BMS deployment, where new charge--discharge data are continuously collected during operation. The results therefore suggest that pre-training on reference ageing trajectories, followed by few-shot fine tuning and rolling-horizon updating, can improve the robustness and transferability of curve-level battery degradation forecasting.

\subsection{Evolution of internal resistance and energy}

\begin{figure}[ht]
    \centering
    \includegraphics[width=\textwidth]{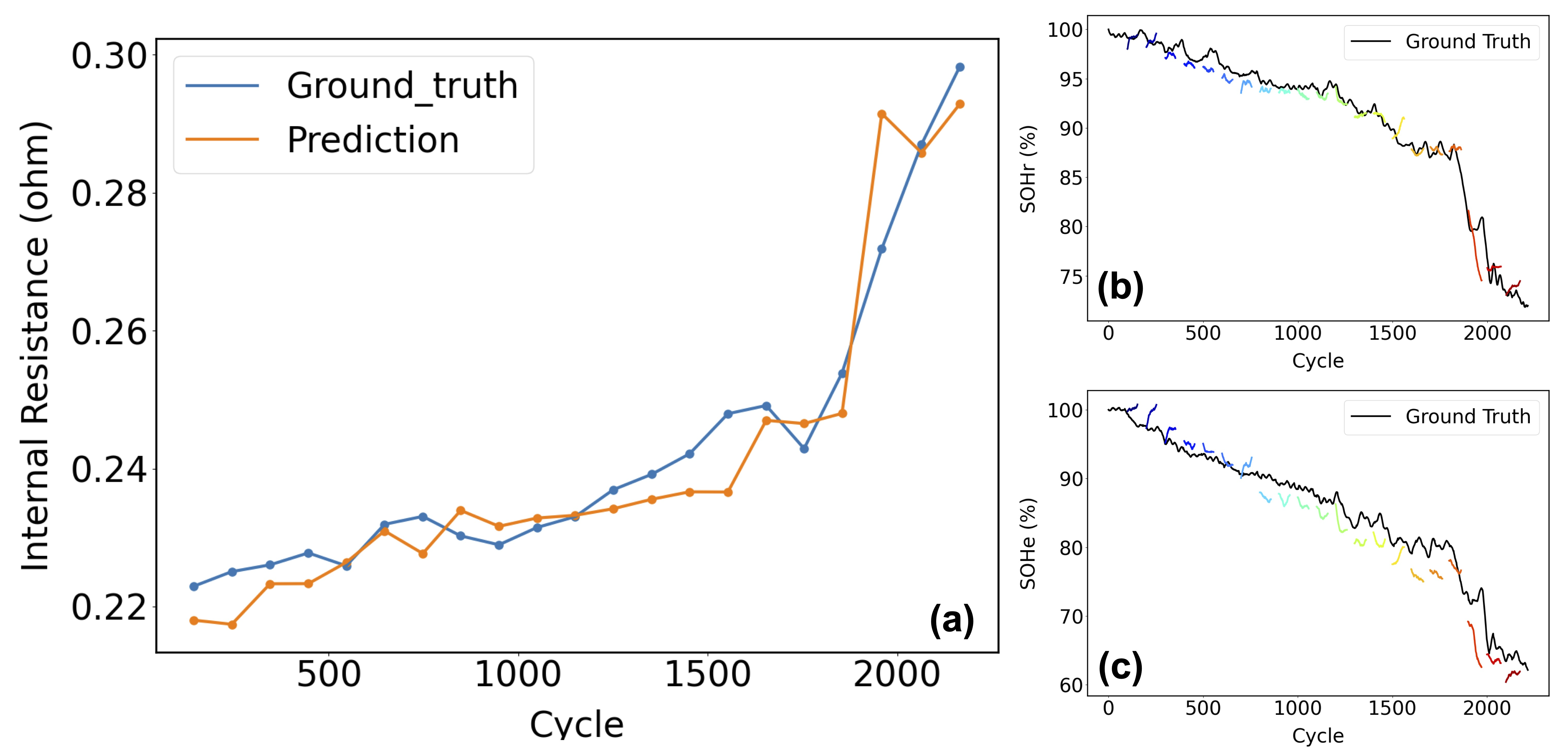}
    \caption{ Evolution of resistance- and energy-based degradation descriptors extracted from the predicted electrochemical profiles. (a) Internal resistance estimated from the relaxation regions of the experimental and predicted voltage--current profiles as a function of cycle number. (b) Resistance-based state of health, SOH$_r$, derived from the evolution of internal resistance and compared between experimental and predicted profiles. (c) Energy-based state of health, SOH$_e$, extracted from the predicted voltage--current profiles and compared with the experimental reference. These results show that the autoregressive curve-level forecasts preserve sufficient electrochemical information to recover both resistance-related ageing and energy-fade trends.}
    \label{resistance}
\end{figure}

Beyond reproducing voltage profiles, an important objective of curve-level forecasting is to determine whether the predicted electrochemical trajectories contain sufficient information to extract physically meaningful degradation descriptors. Figure~\figref{resistance} evaluates this capability through resistance- and energy-based ageing indicators, namely the internal resistance, the resistance-based state of health SOH$_r$, and the energy-based state of health SOH$_e$.

The internal resistance was estimated from the voltage relaxation response following current interruption. As shown in Figure~\figref{resistance}a, the resistance extracted from the predicted profiles follows the same overall trend as the experimental estimate, with a gradual increase during ageing and a stronger rise at later ageing stages. This behavior is consistent with the progressive increase in polarization, kinetic limitations, and transport resistance during cell degradation. Importantly, resistance was not directly used as a supervised target during training. It was extracted a posteriori from the predicted voltage-current profiles, indicating that the model preserves relevant relaxation dynamics in addition to reproducing the global charge-discharge profile shape.

The resistance-based state of health, SOH$_r$, provides a complementary degradation descriptor to capacity- or energy-based indicators. While capacity and energy fade describe the loss of usable storage capability, SOH$_r$ reflects the increase in internal losses and the associated reduction in power capability. The agreement between experimental and predicted SOH$_r$ therefore shows that the autoregressive model captures not only the evolution of the voltage profile, but also the degradation-related changes in cell impedance.

The energy-based state of health, SOH$_e$, was extracted from the energy associated with the predicted voltage--current profiles. Compared with SOH$_r$, SOH$_e$ is more directly related to the usable energy delivered by the cell and therefore provides a global descriptor of performance fade. The predicted SOH$_e$ trajectory follows the experimental trend, confirming that the generated profiles retain sufficient information to estimate energy loss over ageing. Deviations become more pronounced at later ageing stages, where degradation accelerates and small errors in voltage, cycle duration, or current-transition timing can propagate during autoregressive rollout.

The joint analysis of SOH$_r$ and SOH$_e$ highlights the advantage of curve-level forecasting over direct scalar SOH regression. Direct regression generally provides a single health indicator, whereas the present approach predicts the full electrochemical trajectory and allows multiple descriptors to be derived from the same forecast. This is important because resistance growth and energy fade do not necessarily evolve at the same rate and may reflect different degradation mechanisms. Recent studies have emphasized that voltage-profile information encodes rich electrochemical signatures that can be exploited for battery diagnosis and materials analysis~\cite{zhong2024deep}, and that electrical measurements can be used to infer degradation-related thermodynamic and kinetic descriptors~\cite{tao2025degradation}. In this context, the present results show that autoregressive forecasting of voltage--current profiles can provide a unified framework for tracking both local profile evolution and global degradation descriptors such as SOH$_r$ and SOH$_e$.

Overall, these results support the physical relevance of the proposed forecasting strategy. The model does not merely predict a scalar health trajectory, but generates electrochemical profiles from which resistance- and energy-based ageing indicators can be extracted. This capability is particularly relevant for battery diagnostics and management, where energy fade and resistance growth provide complementary information on remaining usability, power capability, and ageing state.

A current limitation of this study is the restricted number of independent ageing trajectories, as the framework was evaluated on two cells cycled under identical conditions. The results should therefore be interpreted as a proof of concept for curve-level autoregressive forecasting, demonstrating its ability to extract physically meaningful degradation descriptors from predicted electrochemical profiles. Broader validation across additional cells, batches, temperatures, current rates, cycling protocols, and chemistries will be required to assess robustness and transferability. This limitation motivates future work toward battery-specific foundation models trained on heterogeneous ageing datasets to improve generalization across cell-to-cell variability and operating conditions.

\section{Conclusion}

In this work, we introduced an autoregressive deep sequence-learning framework for long-horizon forecasting of sodium-ion battery charge-discharge profile evolution and degradation. Instead of directly predicting a single scalar health indicator, the proposed approach forecasts multivariate electrochemical trajectories composed of voltage, current, and cycle index. This curve-level formulation enables the extraction of several degradation descriptors from the same predicted profiles, including capacity-based SOH$_c$, resistance-based SOH$_r$, energy-based SOH$_e$, and internal-resistance evolution.

Two architectures were investigated: a stacked recurrent model, Batt-LSTM, and a hybrid convolutional-recurrent model, Batt-LSTM-C. Both models were able to reproduce the local evolution of voltage profiles, including plateau deformation, relaxation regions, and polarization changes. However, standard sequence-to-sequence training led to error accumulation during long autoregressive rollout, resulting in temporal drift and degradation-trajectory saturation. To address this limitation, scheduled sampling was introduced to reduce the mismatch between training and inference by progressively exposing the models to their own predictions during training. This strategy substantially improved long-horizon stability, extending the stable forecasting horizon from a few tens of cycles to as many as 600 cycles while preserving local waveform fidelity. In the benchmark, Batt-LSTM-SS also achieved the lowest voltage MAE at all evaluated horizons from 5000 to 20\,000 time steps.

The analysis further showed that curve-level forecasting provides richer diagnostic information than direct scalar SOH regression. The predicted electrochemical profiles were successfully converted into SOH$_c$, SOH$_r$, and SOH$_e$ trajectories, demonstrating that the model preserves information related not only to capacity fade, but also to resistance growth and energy loss. Epistemic uncertainty estimation using Monte Carlo dropout and deep ensembles provided estimates of prediction spread at both local voltage-profile and global degradation-trajectory scales. In addition, transfer learning on a second ageing dataset showed that few-shot fine tuning combined with rolling-horizon recalibration can improve model adaptability to a new cell, reducing the SOH$_c$ prediction error to 3.844 percentage points for a 200-cycle update horizon.

These results demonstrate the relevance of autoregressive curve-level forecasting as a data-driven framework for battery ageing analysis and predictive diagnostics. The approach provides a physically interpretable link between raw electrochemical cycling profiles and multiple degradation descriptors, making it suitable for future integration into adaptive battery management systems. More broadly, this work represents a step toward battery-specific foundation models capable of learning transferable representations from large collections of cycling trajectories across chemistries, protocols, temperatures, and usage conditions. Future developments should therefore move beyond cell-specific recurrent models toward foundation models for batteries, combining self-supervised pre-training, multimodal electrochemical data, physics-informed constraints, and uncertainty-aware adaptation. Such models could provide a unified predictive layer for SOH$_c$, SOH$_r$, SOH$_e$, RUL, anomaly detection, and degradation-mechanism identification, enabling more robust, generalizable, and trustworthy battery intelligence.

\section{Statement}
\subsection{Data and code availability}
The code used to train and evaluate the models will be made available upon reasonable request. 
The ageing datasets are subject to restrictions due to industrial ownership by TIAMAT and can be shared only with permission from the data provider.

\subsection{Author contributions}

X.Z. and F.M. contributed equally to this work.
X.Z. contributed to code refinement, re-running forecasting
experiments, comparative analysis of inference results,
benchmarking against classical machine-learning models,
and the critical review, revision, and editing of the
entire manuscript.
F.M. contributed to the conceptualization of the study,
data preparation and curation, software development,
model implementation, formal analysis, validation,
visualization, investigation, and writing of the original draft.
M.A. contributed to code review, debugging, and validation
of the computational workflow.
B.J. contributed to the internal-resistance estimation
procedure and the analysis of resistance-based
degradation descriptors.
R.L. contributed to computational infrastructure,
IT support, and data-management support.
A.D. contributed to the conceptualization of the study,
supervision, resources, project administration,
interpretation of the results, and review and editing
of the manuscript.

\subsection{Declaration of competing interest}
The authors declare that they have no known competing financial interests or personal relationships that could have appeared to influence the work reported in this paper.

\subsection{Acknowledgements}
This work was supported by the French Agence Nationale de la Recherche (ANR) with the Dynamobat project number ANR 22-CE42-0025. The authors would like to sincerely thank Mathieu Morcrette and Asmae El Mejdoub from TIAMAT for their valuable discussions and for granting access to the battery ageing data that supported this study.

\subsection{Supporting Information}

The Supporting Information provides the min--max normalization formula (Equation~S1), pseudocode for the Batt-LSTM and Batt-LSTM-C architectures (Algorithms~S1 and S2), the autoregressive prediction procedure (Algorithm~S3), and the DTW calculation following the \texttt{tslearn} implementation (Algorithm~S4). It also includes the effect of target-window size on prediction quality and temporal drift (Figure~S1), local and global uncertainty estimates (Figure~S2), and the tested hyperparameters (Table~S1).

\printbibliography

\newpage
\maketitle
\thispagestyle{empty}

\begin{center}
    \textbf{\huge Supporting Information}
\end{center}

This Supporting Information provides the normalization procedure, model architectures, autoregressive prediction algorithm, DTW calculation, hyperparameter analysis, and uncertainty estimates accompanying the main manuscript. Equations, algorithms, figures, and tables are numbered independently with the prefix S.

\section{Data normalization}

Voltage, current, and cycle index were normalized separately using min--max scaling:
\begin{equation}
    \operatorname{MinMax}(X)=\frac{X-X_{\min}}{X_{\max}-X_{\min}}.
    \label{min_max}
\end{equation}
Here, $X_{\min}$ and $X_{\max}$ denote the minimum and maximum of the corresponding channel in the training set. The same parameters were applied to the validation, test, and transfer datasets, as described in the main manuscript.

\section{Model architectures}

Algorithms~\ref{alg:batt_lstm} and~\ref{alg:batt_lstm_c} summarize the Batt-LSTM and Batt-LSTM-C architectures. In both algorithms, $L_s$ denotes the number of predicted channels and $T_s$ the target-window length, corresponding to $C$ and $L_{\mathrm{out}}$ in the main manuscript. The selected models use $L_s=3$ channels (voltage, current, and cycle index) and $T_s=10$ time steps.

\subsection{Batt-LSTM}

Batt-LSTM combines three LSTM layers with a dense prediction head. The final output is reshaped into a multivariate sequence of $T_s$ time steps.

\begin{algorithm}[!htbp]
\fontsize{9}{10}\selectfont
\caption{Batt-LSTM architecture}
\label{alg:batt_lstm}
\begin{algorithmic}[1]
\Function{LSTM-Model}{$X, L_s, T_s$}
    \State \textbf{Input:} $X$ (input sequence), $L_s$ (number of channels), $T_s$ (target-window length)
    \State \textbf{Output:} $y \in \mathbb{R}^{T_s\times L_s}$ (predicted sequence)

    \State $h_1, c_1 \gets \text{LSTM}(X, \text{units}=256, \text{activation}=\tanh)$
    \State $h_1 \gets \text{Dropout}(h_1, \text{rate}=0.1)$

    \State $h_2, c_2 \gets \text{LSTM}(h_1, \text{units}=128, \text{activation}=\tanh)$
    \State $h_2 \gets \text{Dropout}(h_2, \text{rate}=0.1)$

    \State $h_3, c_3 \gets \text{LSTM}(h_2, \text{units}=64, \text{activation}=\tanh)$
    \State $h_3 \gets \text{Dropout}(h_3, \text{rate}=0.1)$

    \State $d_1 \gets \text{Dense}(h_3, \text{units}=32, \text{activation}=\text{ReLU})$

    \State $y \gets \text{Dense}(d_1, \text{units}=T_s \times L_s, \text{activation}=\text{sigmoid})$

    \State $y \gets \text{Reshape}(y, \text{shape}=(T_s, L_s))$

    \State \Return $y$
\EndFunction
\end{algorithmic}
\end{algorithm}

\subsection{Batt-LSTM-C}

Batt-LSTM-C uses convolutional and pooling layers to extract local temporal features before recurrent processing. Its prediction head produces the same output shape as Batt-LSTM.

\begin{algorithm}[!htbp]
\fontsize{9}{10}\selectfont
\caption{Batt-LSTM-C architecture}
\label{alg:batt_lstm_c}
\begin{algorithmic}[1]
\Function{CNN--LSTM--Model}{$X, L_s, T_s$}
    \State \textbf{Input:} $X$ (input sequence), $L_s$ (number of channels), $T_s$ (target-window length)
    \State \textbf{Output:} $y \in \mathbb{R}^{T_s\times L_s}$ (predicted sequence)

    \State $h_1 \gets \text{Convolution1d}(X,\ \text{filters}=128,\ \text{kernel}=3,$
    \Statex \hspace{\algorithmicindent}$\qquad \text{strides}=1,\ \text{padding}=\text{same},\ \text{activation}=\text{ReLU})$
    \State $h_1 \gets \text{MaximumPooling1d}(h_1, \text{pool\_size}=2, \text{strides}=2)$
    \State $h_1 \gets \text{Dropout}(h_1, \text{rate}=0.1)$

    \State $h_2 \gets \text{Convolution1d}(h_1,\ \text{filters}=64,\ \text{kernel}=3,$
    \Statex \hspace{\algorithmicindent}$\qquad \text{strides}=1,\ \text{padding}=\text{same},\ \text{activation}=\text{ReLU})$
    \State $h_2 \gets \text{MaximumPooling1d}(h_2, \text{pool\_size}=2, \text{strides}=2)$
    \State $h_2 \gets \text{Dropout}(h_2, \text{rate}=0.1)$

    \State $h_3, c_3 \gets \text{LSTM}(h_2, \text{units}=64, \text{activation}=\tanh)$
    \State $h_3 \gets \text{Dropout}(h_3, \text{rate}=0.1)$

    \State $h_4, c_4 \gets \text{LSTM}(h_3, \text{units}=128, \text{activation}=\tanh)$
    \State $h_4 \gets \text{Dropout}(h_4, \text{rate}=0.1)$

    \State $d_1 \gets \text{Dense}(h_4, \text{units}=64, \text{activation}=\text{ReLU})$

    \State $y \gets \text{Dense}(d_1, \text{units}=T_s \times L_s, \text{activation}=\text{sigmoid})$

    \State $y \gets \text{Reshape}(y, \text{shape}=(T_s, L_s))$

    \State \Return $y$
\EndFunction
\end{algorithmic}
\end{algorithm}

\section{Autoregressive prediction}

Algorithm~\ref{alg:autoregressive} generates a long forecast by recursively feeding each predicted segment back into the model. The initial measured window is processed to produce the first prediction segment and update the recurrent states. Subsequent model calls use the predicted segments together with the states carried over from the preceding call, except when the states are periodically reinitialized.

\begin{algorithm}[!htbp]
\fontsize{9}{10}\selectfont
\caption{Autoregressive prediction with periodic state reinitialization}
\label{alg:autoregressive}
\begin{algorithmic}[1]
\Function{AutoregressivePrediction}{$X, L_s, T_s, N$}
    \State \textbf{Input:} $X$ (initial input window), $L_s$ (number of channels),
    \Statex \hspace{\algorithmicindent}$T_s$ (target-window length), $N \geq 1$ (number of prediction blocks)
    \State \textbf{Output:} $\text{result} \in \mathbb{R}^{(N T_s)\times L_s}$
    \State $R \gets 2000$ \Comment{State-reset interval in predicted time steps}
    \State \textbf{Require:} $R \bmod T_s = 0$
    \State $\text{result} \gets \text{array of shape }(N T_s,L_s)$
    \State $z \gets X$
    \State $\mathcal{S} \gets \text{None}$
    \For{$i \gets 0$ \textbf{to} $N-1$}
        \If{$i > 0$ \textbf{and} $(iT_s) \bmod R = 0$}
            \State $\mathcal{S} \gets \text{None}$ \Comment{Reinitialize all hidden and cell states}
        \EndIf
        \State $(y,\mathcal{S}) \gets \text{model}(z,\text{state}=\mathcal{S})$
        \State $\text{result}[iT_s:(i+1)T_s,\,:] \gets y$
        \State $z \gets y$
    \EndFor
    \State \Return $\text{result}$
\EndFunction
\end{algorithmic}
\end{algorithm}

The model predicts $N \times T_s$ time steps. The hidden and cell states, denoted $\mathcal{S}$, are passed between calls and reinitialized after every 2000 predicted time steps. This state reset does not introduce new measured inputs. The reset interval must be an integer multiple of $T_s$. Array slices use zero-based, end-exclusive indexing.

\section{Dynamic time warping}

Algorithm~\ref{alg:dtw} describes the DTW calculation used in the benchmark. DTW is computed from normalized voltage sequences.

\begin{algorithm}[!htbp]
\fontsize{9}{10}\selectfont
\caption{Dynamic time warping with the \texttt{tslearn} Sakoe--Chiba constraint}
\label{alg:dtw}
\begin{algorithmic}[1]
\Function{DTWDistance}{$s[1..n],t[1..m],w$}
    \State \textbf{Input:} $s,t$ (voltage sequences in V), $w\geq 0$ (integer radius)
    \State \textbf{Output:} normalized-voltage DTW distance
    \State $a \gets s/(2.25\,\mathrm{V})$, $b \gets t/(2.25\,\mathrm{V})$
    \If{$w=0$ \textbf{and} $n=m$}
        \State \Return $\sqrt{\sum_{i=1}^{n}(a_i-b_i)^2}$
    \EndIf
    \State $D \gets \text{array of shape }(n+1,m+1)\text{ filled with }\infty$
    \State $D[0,0] \gets 0$
    \For{$i \gets 1$ \textbf{to} $n$}
        \State $j_{\min} \gets \max\{1,\,i-w-\max(0,n-m)\}$
        \State $j_{\max} \gets \min\{m,\,i+w+\max(0,m-n)\}$
        \For{$j \gets j_{\min}$ \textbf{to} $j_{\max}$}
            \State $\text{cost} \gets (a_i-b_j)^2$
            \State $D[i,j] \gets \text{cost}+\min\{D[i-1,j],D[i,j-1],D[i-1,j-1]\}$
        \EndFor
    \EndFor
    \State \Return $\sqrt{D[n,m]}$
\EndFunction
\end{algorithmic}
\end{algorithm}

A Sakoe-Chiba radius of $w=5$ was used for global DTW and
a radius of $w=0$ for local DTW. Global DTW was computed
over the full forecast and averaged equally over all samples.
Local DTW was computed over matched complete cycles, from
charge start to discharge end, and averaged first over cycles
within each sample and then across samples.

\clearpage
\section{Hyperparameter analysis}

Table~\ref{table:hyperparameter} lists the tested hyperparameter values. Input-window length, target-window length, and stride are expressed in time steps. The number of input channels specifies how many variables are supplied to the model. The selected input and target lengths were 800 and 10 time steps, respectively.

\begin{table}[!htbp]
\centering
\small
\setlength{\tabcolsep}{5pt}
\renewcommand{\arraystretch}{1.15}

\caption{Hyperparameters considered for tuning and their corresponding tested values.}
\label{table:hyperparameter}

\begin{tabularx}{\textwidth}{
    @{}
    >{\raggedright\arraybackslash}p{2.5cm}
    >{\raggedright\arraybackslash}p{3.5cm}
    >{\raggedright\arraybackslash}X
    @{}
}
\toprule
\textbf{Category} & \textbf{Hyperparameter} & \textbf{Tested values} \\
\midrule

Model structure
& Number of LSTM layers
& 1, 2, 3 \\

& Number of dense layers
& 1, 2, 3 \\

& Number of dropout layers
& 1, 2, 3 \\

\addlinespace[3pt]
\midrule

Layer settings
& Number of LSTM units
& 16, 32, 64, 128, 256, 512 \\

& Number of dense units
& 3, 30, 32, 64, 128 \\

& Dropout rate
& 0.05, 0.10, 0.20 \\

\addlinespace[3pt]
\midrule

Data
& Number of input channels
& 1, 2, 3, 4 \\

& Input window length ($L_{\mathrm{in}}$)
& 5, 10, 15, 20, 30, 50, 100, 125, 150, 175, 200,
250, 300, 400, 500, 600, 750, 800, 1000, 1800,
2400, 3000, 4000, 5000, 10\,000 \\

& Forecast window length ($L_{\mathrm{out}}$)
& 1, 5, 10, 25, 50, 100, 200, 300, 500, 1000 \\

& Sliding-window stride ($s$)
& 1, 10 \\

\bottomrule
\end{tabularx}
\end{table}

\begin{figure}[!htbp]
    \centering
    \includegraphics[width=\textwidth]{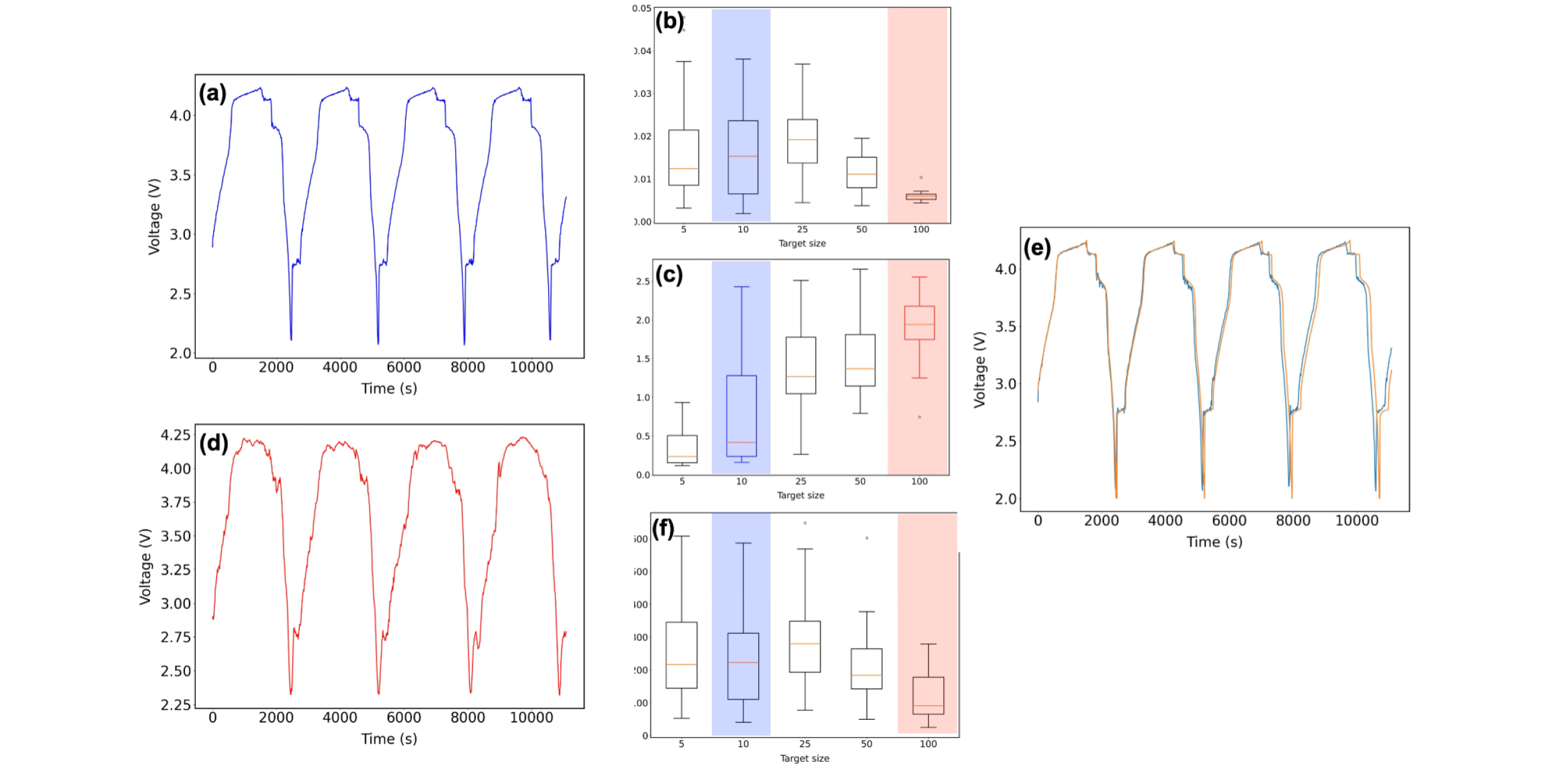}
    \caption{\textbf{a)} Voltage versus time for a target size of 10. \textbf{b)} Boxplots of MSE across parameter configurations at each fixed target size. \textbf{c)} Boxplots of cycle-wise DTW across parameter configurations at each fixed target size. \textbf{d)} Voltage versus time for a target size of 100. \textbf{e)} Example of temporal drift between the ground truth and the prediction. \textbf{f)} Boxplots of average drift across parameter configurations at each fixed target size.}
    \label{fig:result}
\end{figure}

Figure~\ref{fig:result} illustrates the effect of target-window length on prediction quality and temporal drift. Larger targets tend to reduce drift but produce noisier voltage profiles. A target length of 10 time steps was selected to balance these effects.

\section{Local and global uncertainty}

\begin{figure}[htbp]
    \centering
    \includegraphics[width=\textwidth]{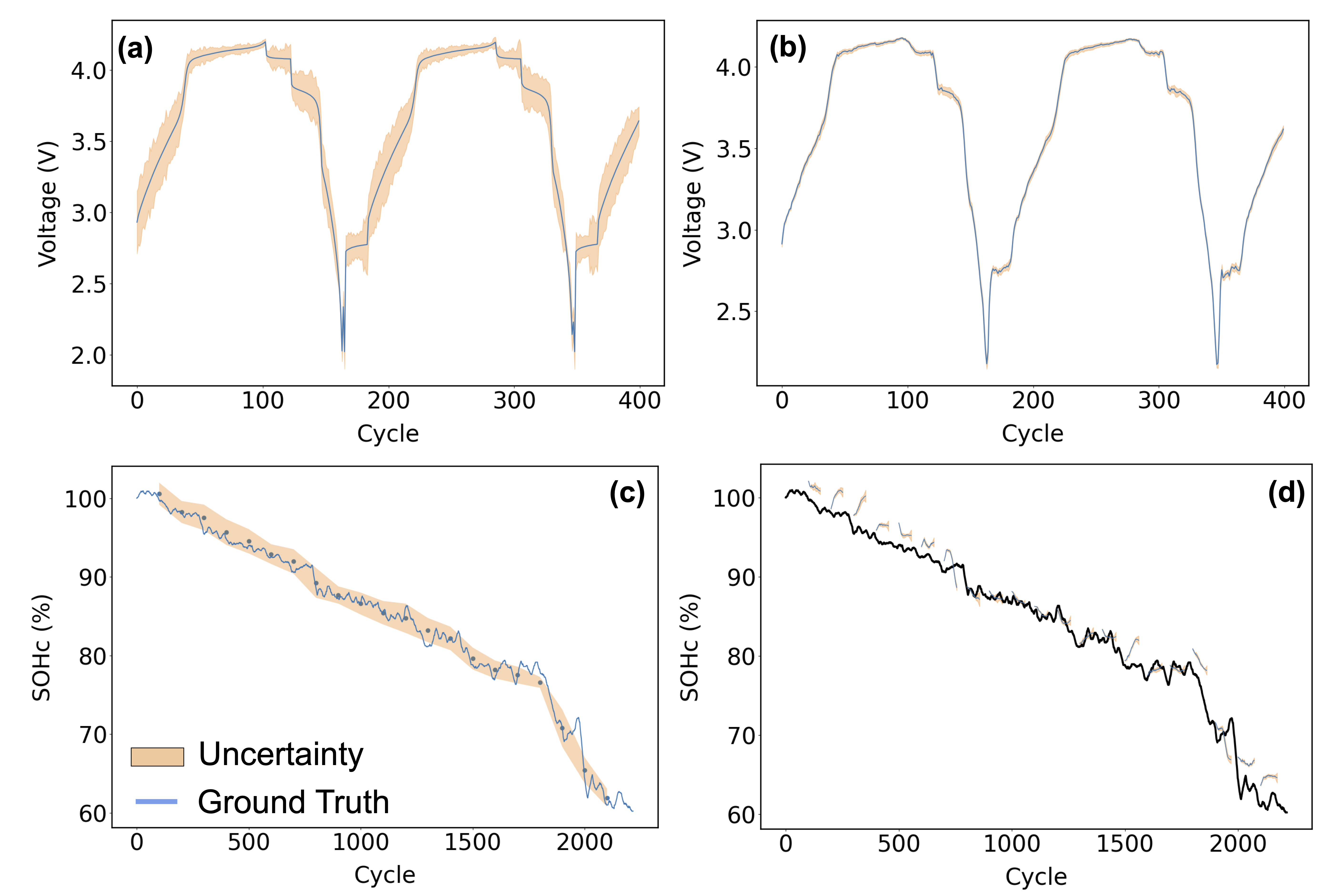}
    \caption{Local and global uncertainty in autoregressive forecasting. (a,b) Voltage-profile predictions with MC-dropout uncertainty estimates for the standard Batt-LSTM and scheduled-sampling models, respectively. (c) SOH$_c$ predictions with uncertainty estimated using independently trained Batt-LSTM models. (d) SOH$_c$ predictions with uncertainty estimated using repeated MC-dropout autoregressive rollouts for the scheduled-sampling model. Shaded regions show the estimated prediction spread.}
    \label{fig:uncertainty}
\end{figure}

Figure~\ref{fig:uncertainty} complements the uncertainty analysis in the main manuscript. Local voltage-profile uncertainty was estimated using 1000 stochastic forward passes per model call, with dropout active at a rate of 0.1. The mean predicted segment was fed back as the input to the next model call, while the pointwise standard deviation quantified the local prediction spread.

Global SOH$_c$ uncertainty was estimated using independently trained models for the standard Batt-LSTM configuration and repeated MC-dropout autoregressive rollouts for the scheduled-sampling configuration. Because these estimators differ, their uncertainty estimates should not be interpreted as a direct comparison of the training strategies alone.

\end{document}